\documentstyle[aps]{revtex}
\begin{document}
\draft
\title{\bf String Fields and the Standard Model}
\author{ T. Dereli}
\address{ Department of Physics,
Middle East Technical University,\\ 06531 Ankara, Turkey\\ 
{\footnotesize tekin@dereli.physics.metu.edu.tr}}
\author{Robin W. Tucker} 
\address{Department of Physics, 
University of Lancaster,\\ Bailrigg, Lancs. LA1 4YB, UK\\
{\footnotesize r.tucker@lancaster.ac.uk}}
\date{9 August 1998 }
\maketitle
\begin{abstract}
\noindent The  Cremmer-Scherk mechanism is  generalised in a non-Abelian context.  
In the presence of the Higgs scalars of the standard model it is argued that fields 
arising from the low energy effective string action may contribute to the mass
generation of the observed vector bosons that mediate the  electroweak interactions
and that future analyses of experimental data should
consider the possibility of string induced radiative corrections
to the Weinberg angle coming from physics beyond the standard model.

\end{abstract}

\vskip 4mm
The standard model of fundamental particle interactions is based on families of charged leptons,
 quarks and scalar fields interacting via gauge fields associated with 
Lie groups [1],[2]. This structure is general enough to accommodate gravitation and it is
a common belief that an underlying geometrical theory associated with some
extended structure in higher dimensions will offer a unified framework
from which the standard model will emerge as an effective low energy approximation.
Although such a unique unified geometrical theory is currently elusive,
there are strong hints that it may manifest itself in a number of different
guises. These alternative representations are said to be related by certain
duality transformations.
They promise to relate different spectra associated with dual coupling regimes
and offer an explanation of the observed masses of the fundamental particles.
At the level of low energy  effective actions, however, current phenomenology relies
on a number of alternative mechanisms for mass generation.
These may be briefly classified as (Kaluza-Klein) harmonic expansions
in dimensional reduction, dynamical symmetry breaking and the Higgs mechanism
based on the existence of fundamental self-coupled scalar fields.
Each mechanism makes particular predictions and experimental guidance
is required for preferring one without prejudice.

In this note we explore the role of the Abelian gauge
 vector potential  $A$, the dilaton $\phi$ and the antisymmetric 3-form   gauge field $H$
that arise in certain low energy effective actions on the Higgs mechanism of the standard model.
In lowest order $H$ is usually interpreted as describing a scalar (axion) field that, together with
the dilaton scalar $\phi$,  modifies   Einsteinian gravitation.
\footnote{The interpretation of a massless axion ${\cal A}$ is based on the field equation $d(e^{-2\phi}\,*dB)=0$ so that locally $dB=e^{2\phi}\,* d{\cal A}$ and hence $d(e^{2\phi}\,*d{\cal A})=0$.} However, we propose here that $H$ couples to vector fiel
ds via 
an interaction with Higgs fields in such a way 
that it contributes to  mass generation
via a spontaneous breakdown of gauge symmetry. The coupling is inspired by an earlier
incarnation of string theory as a dual model for the strong interactions
and the recognition that sectors associated with ``Reggeons'' and ``Pomerons'' are
``dual''. The  gauge potential $A$  is analogous to a 
Reggeon state in the hadron model and the Abelian gauge symmetry in the action
is inherited from a representation of the Virasoro symmetry. The 
antisymmetric tensor field described here in terms of the 2-form $B$, where to lowest order
 $H=dB$, corresponds
to a  state in the Pomeron sector of the dual model.
In 1974, Cremmer and Scherk suggested on the basis of a 1-loop summation that both these fields
could gain mass by radiative corrections [3],[4]. Furthermore they gave a gauge invariant effective action
that maintained a duality between the Reggeon and Pomeron descriptions.

In the context of low energy effective string theory, the mechanism arises
from the effective action density 4-form
\begin{eqnarray}
\Lambda[{\bf g}, \phi, A, B] &=& \kappa {\cal R}*1 - \frac{(2\alpha-3)}{4}
d\phi \wedge *d\phi \nonumber \\
& & + \frac{1}{2} e^{-2\phi} dB \wedge *dB
+ \frac{1}{2} e^{-2\phi} dA \wedge *dA
+ \lambda A \wedge dB
\end{eqnarray}
where $A$ is a 
1-form, $B$ a 2-form, $\phi$ the dilaton 0-form
on spacetime $M$ with a metric ${\bf g}$, curvature scalar $\cal R$
 and associated Hodge map $*$. The action $\int_{M} \Lambda$
is invariant under the Abelian gauge symmetries
$A \rightarrow A + df_0$, $B \rightarrow B+ df_1$ where $f_0$ and $f_1$
are arbitrary 0 and 1 forms, respectively.
In terms of the Weyl scaled metric $\hat{{\bf g}} =e^{-\phi}{\bf g}$, with the 
corresponding Hodge map $\star$ and curvature scalar $\hat{{\cal R}}$,
the effective action may be written
\begin{equation}
\int_{M} \{e^{-\phi} ( \kappa \hat{{\cal R}}\star 1
- \frac{\alpha}{2}d\phi \wedge \star d\phi
+ \frac{1}{2}dB \wedge \star dB)
+ \frac{1}{2} e^{-2\phi} dA \wedge \star dA
+ \lambda A \wedge dB \}.
\end{equation}
In this form one can make contact with  low energy effective
axi-dilaton gravity when $\lambda = 0$ [5],[6].
Having established this link, we concentrate next on the dynamics of the
field system $\{\phi, B, A\}$ in a flat Minkowski background
(the implications for the gravitational sector will be discussed elsewhere):
\begin{equation}
d({e^{-2\phi}} *dA) + \lambda dB = 0,
\end{equation}
\begin{equation}
d({e^{-2\phi}} * dB) - \lambda dA = 0,
\end{equation}
\begin{equation}
d*d\phi = \frac{2}{(2\alpha-3)} e^{-2\phi}(dB \wedge *dB
 + dA \wedge *dA).
\end{equation}
Since $M$ is topologically trivial (3) and (4) imply
\begin{equation}
d\tilde{A}= \lambda e^{2\phi} *\tilde{B}, 
\end{equation}
\begin{equation}
d\tilde{B}= \lambda e^{2\phi} *\tilde{A}, 
\end{equation}
in terms of the variables $\tilde{A} = A - \frac{1}{\lambda}df_0$,
$\tilde{B} = B - \frac{1}{\lambda}df_1$ in the gauge equivalence classes
$[A]$ and $[B]$, respectively.
One may  fix  gauges by taking solutions with particular $f_0$ and $f_1$.
Remarkably the entire theory can be described in terms of either the fields
$\{\phi, \tilde{A}\}$ or $\{\phi, \tilde{B}\}$. Each  description
refers to a  different dual sector 
of the same theory. Moreover, in terms of the $\{\phi, \tilde{A}\}$ description the
theory admits a vector field satisfying the dilaton-Proca system:
\begin{equation}
d(e^{-2\phi} *d\tilde{A}) + {\lambda}^{2} e^{2\phi} *\tilde{A} = 0,
\end{equation}
\begin{equation}
d*d\phi = - \frac{2\lambda^2}{(2\alpha-3)} e^{2\phi}\tilde{A} \wedge *\tilde{A}
+ \frac{2}{(2\alpha-3)} e^{-2\phi} d\tilde{A} \wedge *d\tilde{A}.
\end{equation}
These equations admit a ``vacuum'' solution $\tilde{A} = 0$, $\phi = \phi_0$ for some
constant $\phi_0$. Linearising about this solution  with
$\tilde{A} = \epsilon A^{(1)}$, $ \phi = \phi_0 + \epsilon \phi^{(1)}$ yields
\begin{equation}
d *d\tilde{A^{(1)}} + {\lambda}^{2} e^{4\phi_0} *\tilde{A^{(1)}} = 0,
\end{equation}
\begin{equation}
d*d\phi^{(1)} = 0,
\end{equation}
showing that the Abelian gauge field acquires a mass $\mu_{0} = \lambda e^{2\phi_0}$.
In a similar way the theory has a dual description in terms of $\{\phi, \tilde{B}\}$:

\begin{equation}
d(e^{-2\phi} *d\tilde{B}) - {\lambda}^{2} e^{2\phi} *\tilde{B} = 0,
\end{equation}
\begin{equation}
d*d\phi = 
- \frac{2 \lambda^2}{(2\alpha-3)} e^{2\phi} \tilde{B} \wedge *\tilde{B}
+ \frac{2}{(2\alpha-3)} e^{-2\phi}d\tilde{B} \wedge *d\tilde{B},
\end{equation}
showing that the Kalb-Ramond 2-form field $\tilde B$ also acquires
a mass $\mu_0$.

\def\bfA{{\bf A}}
\def\bfD{{\bf D}}
\def\DD{{\cal D}}
\def\bfPhi{{\bf \Phi}}
The role of the fields $\{\phi, \tilde{A}, \tilde{B}\}$
is elusive in the standard model. However, guided by the $\tilde{A} \leftrightarrow 
\tilde{B}$ duality in the presence of the Cremmer-Scherk mechanism
we propose that the Abelian gauge potential $A$ be identified with the weak
hypercharge gauge potential in the electroweak $SU(2) \times U(1)$ gauge group
and that the Cremmer-Scherk  mass generation mechanism be generalised with the aid of
the standard Higgs multiplet.
The low energy effective action density (for the bosonic sector of the standard electroweak model)
 now includes the Higgs isospinor $\bfPhi$ and the $SU(2)$
gauge potential $\bfA$: 
\begin{eqnarray}
\Lambda[\phi, A, B, \bfA, \Phi] =&-& \frac{(2\alpha-3)}{4}
d\phi \wedge *d\phi
+ \frac{1}{2} e^{-2\phi} dB \wedge *dB  \\
&+& \frac{1}{2} e^{-2\phi} \{ dA \wedge *dA + Tr({\bf F} \wedge *{\bf F}) \} \nonumber \\
&+& \frac{1}{2} (\DD \bfPhi)^{\dagger} \wedge *\DD \bfPhi + i \lambda \bfPhi^{\dagger} 
\DD \bfPhi \wedge dB + V(\vert \bfPhi \vert) *1 \nonumber
\end{eqnarray}
where ${\bf F} = d\bfA + g \bfA \wedge \bfA$ with $\bfA$ an $SU(2)$ Lie algebra 
(with basis ${\bf T}_j$) valued 1-form and
$\DD \bfPhi = d \bfPhi + g \bfA \bfPhi + i \frac{g^{\prime}}{2} A \bfPhi$ with
$\bfA = A_{j}\frac{{\bf t}_j}{2i}$ in terms of the Pauli matrices ${\bf t}_j$. 
The dynamics of the fields $\{ \bfA, A \}$ arise from the field equations
\begin{equation}
\bfD(e^{-2\phi} *{\bf F}) + g \lambda ({\bfPhi}^{\dagger} \frac{{\bf t}_j}{2}\bfPhi) {\bf T}_{j} dB
+ \frac{g}{2} *({(\DD\bfPhi)}^{\dagger}\frac{{\bf t}_j}{2i}\bfPhi - \bfPhi^{\dagger}\frac{{\bf t}_j}{2i} \DD\bfPhi) {\bf T}_j =0, 
\end{equation}
\begin{equation}
d(e^{-2\phi}*dA) -\frac{g^{\prime}}{2}\lambda {\bfPhi}^{\dagger}\bfPhi dB 
-i\frac{g^\prime}{4} *( {\bfPhi}^{\dagger}\DD\bfPhi -(\DD\bfPhi)^{\dagger}\bfPhi) = 0
\end{equation}
where $\bfD = d + g\bfA$ with $\bfA$ in the adjoint representation of $SU(2)$.
In the Higgs vacuum
$ \bfPhi = (v , 0)^T$, the new interaction 
$i \lambda {\bfPhi}^{\dagger} \DD \bfPhi \wedge dB$ becomes
 $\frac{\lambda {\vert v \vert}^2}{2} (g A_{3} - g^{\prime}A) \wedge dB$
(cf. the interaction in (1)) and one expects that the mass generation for the vector fields will be modified.
This is indeed the case. Expanding about the vacuum, $\bfPhi = (v , 0)^T$, $\phi = \phi_0$,
with $dB=i\,\lambda\, e^{2\phi}\,*\,(\bfPhi^\dagger \DD\bfPhi)$, the mass eigenstates are given by
the equations 
\begin{equation}
d(e^{-2\phi_0} *dA_1) + \frac{g^2}{4}{\vert v \vert}^{2} *A_1 = 0,
\end{equation}
\begin{equation}
d(e^{-2\phi_0} *dA_2) + \frac{g^2}{4}{\vert v \vert}^{2} *A_2 = 0, 
\end{equation}
\begin{equation}
d(e^{-2\phi_0} *dA_3) + \frac{\lambda^2}{4}e^{2\phi_0} g{\vert v \vert}^{4} *(gA_3 -g^{\prime}A)
+ \frac{g^2}{4}{\vert v \vert}^{2} *A_3 - \frac{gg^\prime}{4}{\vert v \vert}^2 *A = 0  
\end{equation}
\begin{equation}
d(e^{-{2\phi}_0}*dA) - \frac{g^\prime}{4} {\vert v \vert}^{2}({\lambda^2}e^{2\phi_0}{\vert v \vert}^{2}+1)
*(g A_{3}-g^{\prime}A) = 0.
\end{equation}
A straightforward diagonalisation of these equations shows that the physically charged 
eigenstates $\{ W^{\pm}, Z^{0}, \gamma \}$ have masses
\begin{eqnarray}
M_{W^{\pm}} &=& e^{\phi_0}\,\frac{g}{2} {\vert v \vert}, \\
M_{Z^0} &=& \frac{ e^{\phi_0}}{2} {\vert v \vert}\sqrt{(g^2+{g^\prime}^2)(1+{\lambda}^{2}e^{2\phi_0}
{\vert v \vert}^{2}) },  \\
M_{\gamma} &=& 0 
\end{eqnarray}  
and  
\begin{equation}
\frac{M_{Z^0}}{M_{W^\pm}} = \sec \theta_{W} \sqrt{1+{\lambda}^{2}e^{2\phi_0}
{\vert v \vert}^{2}}
\end{equation}
depends explicitly on the strength of the effective interaction of the
Higgs field with the low energy  string fields. 

We have argued that the Cremmer-Scherk mechanism can be generalised 
in the presence of the Higgs fields of the standard model and that the fields 
arising from the low energy effective string action may contribute to the mass
generation of the observed vector bosons that mediate the  electroweak interactions.
This result suggests that future analyses of experimental data should
consider the possibility of string induced radiative corrections
to the Weinberg angle coming from physics beyond the standard model [7].

\vskip 1cm
The authors are grateful to TUBITAK for the support of this research.
 RWT is also grateful to the Department of Physics, Middle East Technical University 
for hospitality.

\newpage
\vskip 8mm
\noindent {\bf References}
\vskip 4mm
\begin{description}
\item{[1]} S. Weinberg, Phys. Rev. Lett. {\bf 19}(1967)1264
\item{[2]} A. Salam, in {\bf Elementary Particle Theory} edited by N. Svartholm
(Almqvist \& Wiksell, 1968) p.367   
\item{[3]} E. Cremmer, J. Scherk, Nucl. Phys. {\bf B72}(1974)117
\item{[4]} M. Kalb, P. Ramond, Phys. Rev. {\bf D9}(1974)2273
\item{[5]} A. Shapere, S. Trivedi, F. Wilczek, Mod. Phys. Lett. {\bf A6}(1991)2677
\item{[6]} T. Dereli, M. \"{O}nder, R. W. Tucker, Class. Q. Grav. {\bf 12}(1995)L25
\item{[7]} P. Langacker, M. Luo, A. K. Mann, Rev. Mod. Phys. {\bf 64}(1992)87
\end{description} 
\end{document}